\documentclass[12pt]{article}
\usepackage{graphicx}
\usepackage{fullpage}
\author{N. Meinshausen, J. Rice and T. Sch\"ucker \\ \textit{University of California, Berkeley, and Universit\'{e} de Provence,  Marseille} \\nicolai@stat.berkeley.edu\\ rice@stat.berkeley.edu\\ schucker@cpt.univ-mrs.fr}
\title{Testing for monotonicity in the Hubble diagram}
\begin{document}
\maketitle

\begin{abstract}
General relativistic kinematics and the cosmological principle alone imply a monotonicity constraint in the Hubble diagram, which we confront to present-day supernova data.
We use the running gradient method of statistical inference by Hall \& Heckman (2000).
We find no significant departure from monotonicity. The method seems well adapted and we recommend its use with future data.
\end{abstract}

\section{Introduction}
The Hubble diagram is one of the main pillars of modern cosmology.
Indeed the accuracy in the measurement of apparent luminosity
$\ell$ in Watt/m$^2$ and redshift $z$ of supernovae has
dramatically increased in the last 20 years and is expected to
continue so in the future. The theoretical understanding of the
Hubble diagram remains a matter of lively discussion. A majority
of view-points however accepts the kinematical assumptions of
general relativity and the cosmological principle. These weak
hypotheses are sufficient to produce a monotonicity constraint in
the Hubble diagram of standard candles: in an open, expanding
universe the apparent lumionosity is a decreasing function of
distance and consequently of redshift. More stringently, the
function $g(z):=(z+1)^2 \ell (z)$ must be decreasing. The two
factors $(z+1)$ come from the energy loss in the photon flux due
to expansion induced redshift and surface increase. In a closed
universe, the apparent luminosity may increase thanks to
gravitational macro-lensing. The function $g(z)$ can go through at
most one minimum as the photon goes over the equator. For higher
$z$ then, $g(z)$ must be increasing. Its smallest observed  value
yields a lower bound for the radius of the universe. Our best
experimental data today are the 137 supernovae of the `Gold'
sample compiled by Riess et al. (2004) \cite{gold} yielding a
lower bound of $1.2 \cdot 10^{26}$ m and a preliminary statistical
analysis sees no sign of non-monotonicity in $g(z)$, Sch\"ucker \&
Tilquin (2006) \cite{andre}.

The origins of order restricted statistical inference go back to
the early 1950's \cite{book1,book2} and the field continues to develop
for natural reasons: many types of problems in the real world are
concerned with monotonic functions. For example, the probability
of a particular response may increase with the treatment level,
the failure rate of a component may increase as it ages.

The purpose of this note is to apply the state of the art in
testing monotonicity of a regression to the Hubble diagram. We
choose the `running gradient' type method by Hall \& Heckmann
(2000) \cite{hh}, and develop a modification suitable to the
closed universe case.

\section{Methodology}

Denote the data by $(z_i, Y_i, \sigma_i)$, $i=1, \ldots , n$,
where the redshifts $z_i$ are indexed in increasing order,  $Y_i
=\log_{10}g(z_i)$ ,  and $\sigma_i$ is the standard deviation of
the measurement $Y_i$.

The basic idea of a family of tests is the following:  for a
region $R(r,s)=[z_{r+1},z_s]$ with $s>r+1$, 
calculate a measure $T(r,s)$ of the departure from
monotonicity of the relation of $Y_i$ and $z_i$, for $z_i \in
R(r,s)$.  The overall test statistic is $T = \sup_{r,
s} T(r,s)$, where the supremum is taken over a
predetermined set of pairs $(r,s)$.  The test thus scans for
deviations from monotonicity over a range of locations and scales.
The p-value for the test is determined by comparing the observed
value of $T$ to the distribution determined by simulation when
monotonicity holds.

There are obviously a variety of constructions of this form; we
use one proposed and analyzed in \cite{hh}. To test the null
hypothesis that $g(z)$ is monotone versus any deviation from
monontonicity the regions are of the form $[z_{r+1}, z_s], ~ 0
\leq r \leq s-m \leq n-m$, where $m \geq 2$ is a parameter to be
specified.  To test versus the alternative that the $g(z)$ is
non-monotone but with a single minimum, the regions are of the
form $[z_{r+1}, z_n]$.  In each region the measure of departure
from monotonicity is the slope of the weighted least squares line
fit to the corresponding $(z,Y)$ pairs divided by its standard
error. The weights are determined by the $\sigma_i$. The overall
test statistic  of the null hypothesis that the relationship is
monotone decreasing versus the unconstrained alternative
hypothesis that is it not is thus
\begin{equation}\label{T-stat}
T = \max_{0 \leq r \leq s-m \leq n-m} T(r,s), 
\end{equation}
where $T(r,s)=\hat{b}(r,s)/ \sigma_{\hat{b}(r,s)}$ with
\begin{eqnarray}
\hat{b}(r,s)& = &\frac{ \sum w_i Y_i (z_i - \bar{z}_w)}{\sum w_i
(z_i -
\bar{z}_w)^2} ,\\
\sigma_{\hat{b}(r,s)} &= & \left( \frac{1}{ \sum w_i(z_i -
\bar{z}_w)^2} \right)^{1/2}.
\end{eqnarray}
Here the summation runs over  $i=r+1, \ldots, s$,  $w_i =
1/\sigma_i^2$ and $\bar{z}_w = \sum w_i z_i/\sum w_i$.  The
constrained test is as above, but with $s=n$.

The significance of the test statistic $T$ is determined by
simulation.  We assume that the errors are Gaussian with mean zero and
variances $\sigma_i^2$.  In \cite{hh} it is shown that simulating
under the model of a constant relationship provides p-values that
are conservative, in the sense that if one were able to simulate
from the true monotonically decreasing relationship, the p-value
thus obtained would be stochastically smaller than the p-value
obtained by simulating from a model in which the relationship was
constant. It is also shown that the results are somewhat robust to
deviations from the Gaussian model.  One thus simulates from a
model in which the $z_i$ are those in the actual data and the
$Y_i$ are Gaussian with means zero and variances $\sigma_i^2$.

\section{Numerical Results}
\begin{figure}
\begin{center}
\includegraphics[width=0.6\textwidth]{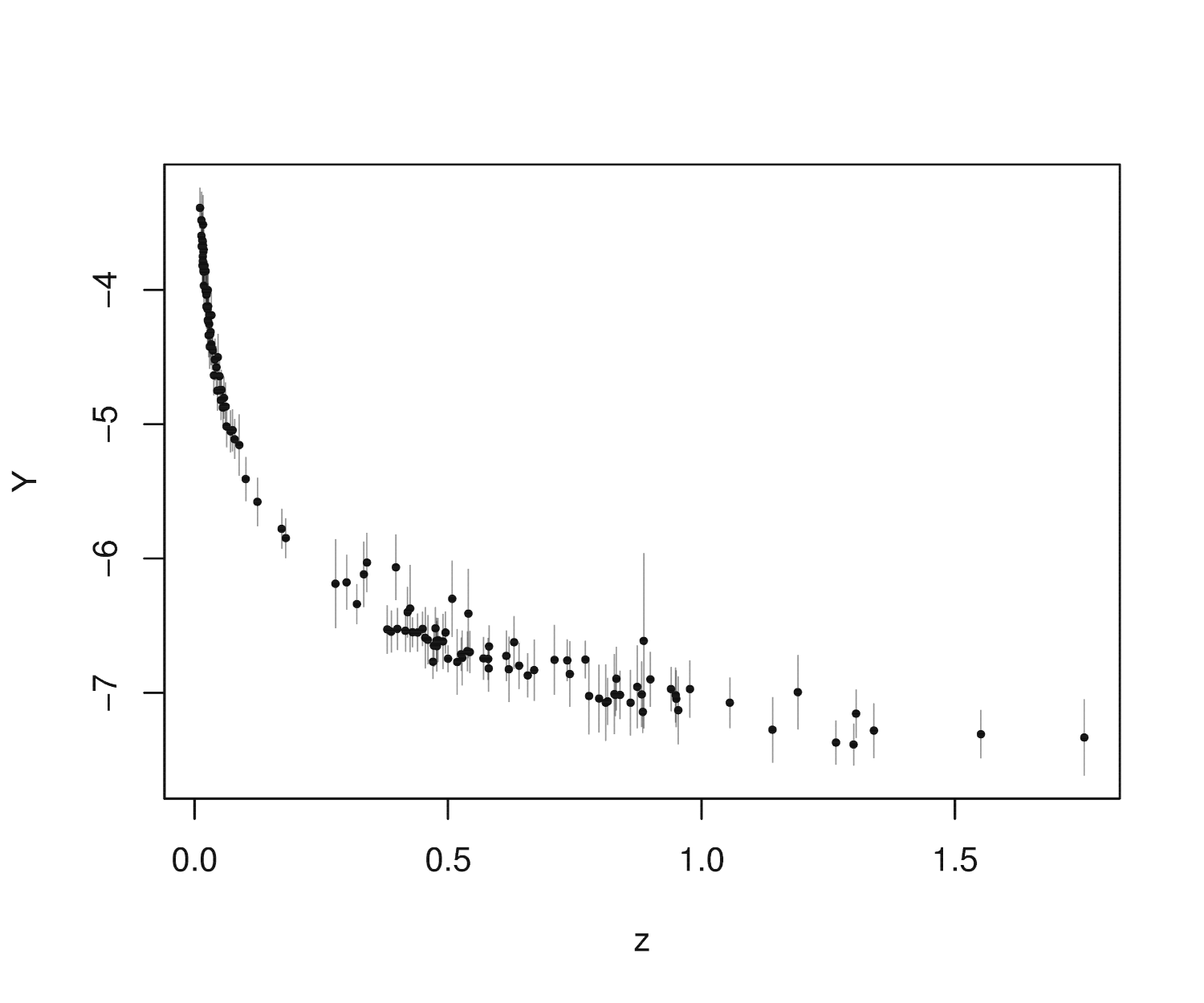}
\caption{ \label{fig} All 137 `Gold' samples of Riess et al.
(2004), showing $Y_i=\log_{10} g(z_i)$ as a function of $z_i$ for
$i=1,\ldots,137$. Error bars show a range $\pm 2 \sigma_i$ for
each observation. }
\end{center}
\end{figure}

\begin{figure}
\begin{center}
\includegraphics[width=0.45\textwidth]{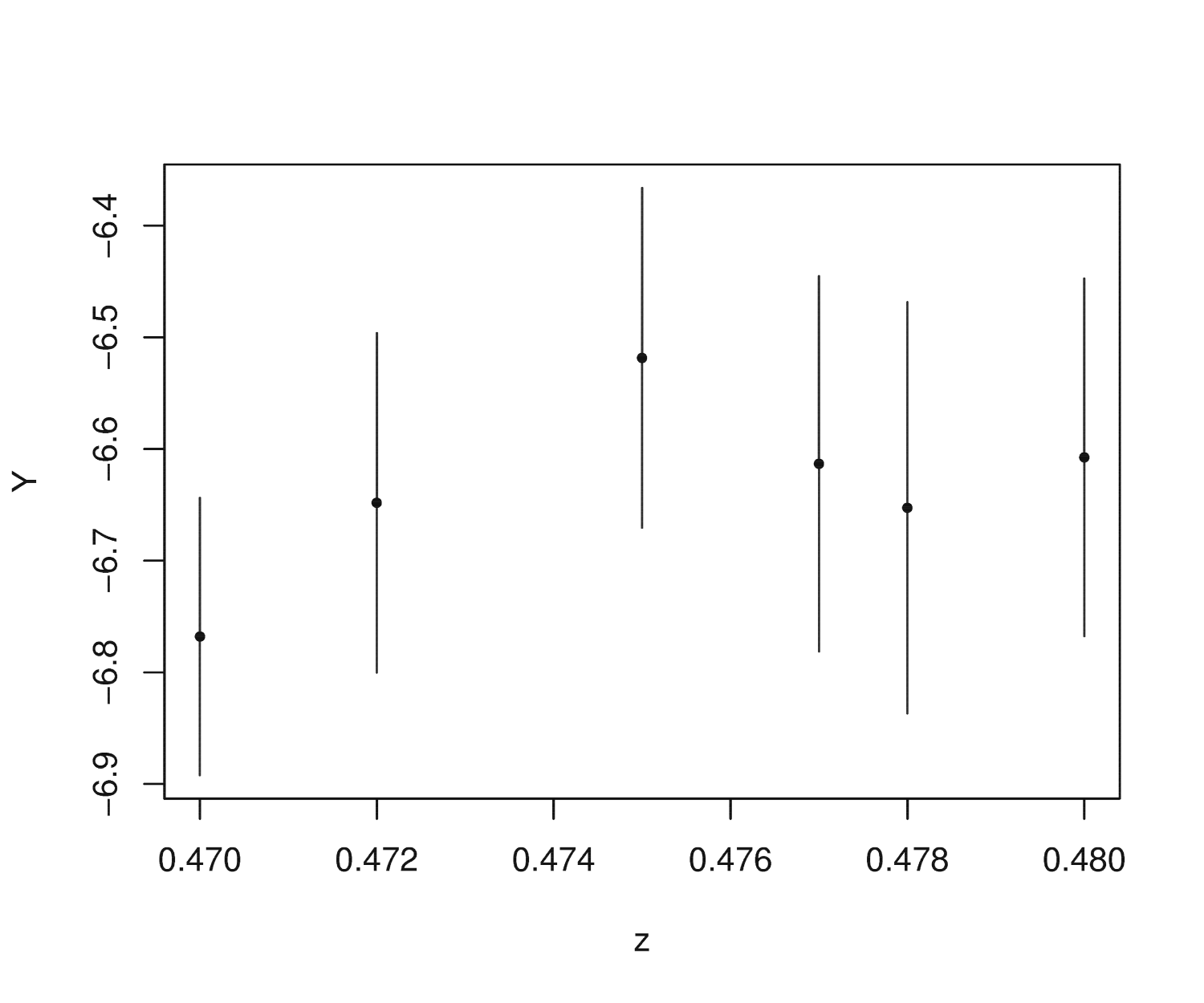}
\includegraphics[width=0.45\textwidth]{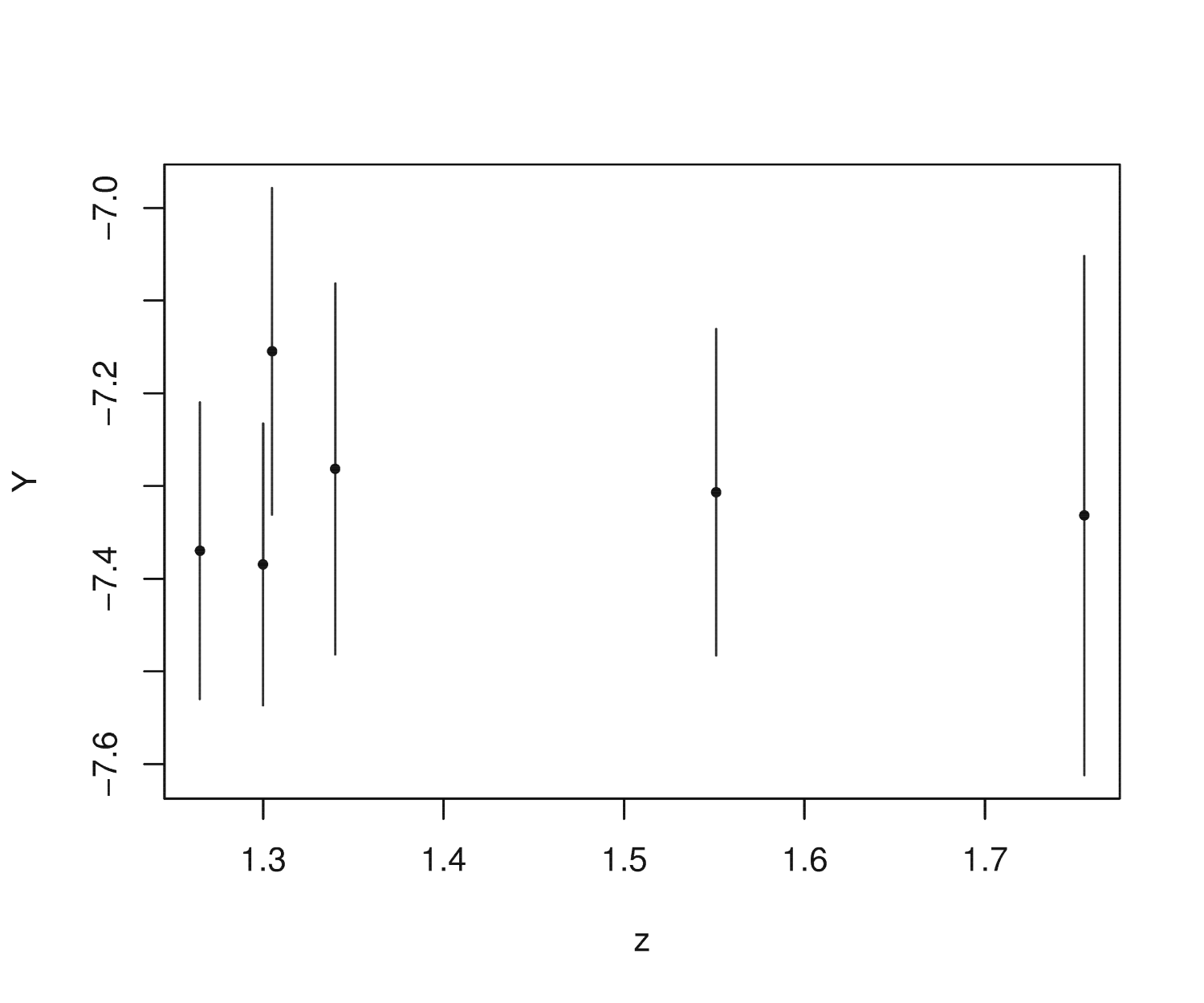}
\caption{ \label{fig2} The regions with the largest deviation from monotonicity for the unconstrained test (left) and constrained test (right).   }
\end{center}
\end{figure}

Assuming approximately normal distributed errors in the
measurement of the distance modulus, the methodology of \cite{hh}
can be applied to the `Gold' sample by {Riess et al.} (2004). The
data are shown in  Figure~\ref{fig}. The 137
observations $i=1,\ldots,137$ consist of triplets $(z_i, Y_i ,
\sigma_i)$, where  $Y_i=\log_{10} g(z_i)$ and $\sigma_i$ is the
standard error of the observation of $Y_i=\log_{10} g(z_i)$. As
$Y_i$ is proportional to $-2 \mu_i/5$, where $\mu_i$ is the
observed distance modulus, the standard error $\sigma_i$ is given
by the standard error of the observation of $\mu_i$ times $2/5$.

We choose a value of $m=6$, so that at least six observations
fall into each considered interval (results are very insensitive
to this choice).

\subsection{No prior}

If the test statistic is calculated over all possible intervals, a
stretch of observations between 0.47 and 0.495 attains the largest
standardized slope of 1.677. This
region is shown in the left panel of Figure~\ref{fig2}.  However, if assessing the
significance of this result by 1000 simulations of the test
statistic under the assumption of a constant $Y(z)$, all 1000
simulations yield a larger test statistic. The observations do
therefore not represent a significant departure from a
monotonically decreasing function $g$. For smaller values of $m$ than the chosen value $m=6$, the region 
with largest standardized slope contains the observations in the leftmost part of the region [0.47,0.495]. The obtained standardized slopes are, as for $m=6$, not even close to being a significant departure from monotonicity.  
 The conclusion are thus insensitive to the choice of the minimal number of observations  $m$ that have to fall within a region over which the standardized slope is calculated. A larger value of $m$ like the chosen $m=6$ has the advantage of being a more robust estimation procedure. If errors of the observations are in effect more heavy-tailed than a Gaussian distribution would suggest, results are more reliable for larger values of $m$, as the effect of an outlier in the data has less effect on the overall result if a larger number of observations are available for each interval over which the standardized slope is calculated.
 The compatibility of the observations with monotonicity and the
location of the window of largest deviation are in agreement with
the preliminary test in \cite{andre}.

\subsection{Assuming a closed universe}

Let us suppose that the universe is closed and that its typical
minimum of $g(z)$ is not hidden behind our horizon. Then
deviations from monotonicity are to be expected for larger
redshifts and it is natural to constrain the test to intervals
including the observation with the largest value of $z=1.755$. The
constrained test would be expected to be more powerful, since the
maximum in (\ref{T-stat}) is taken over a smaller, targeted, set.
The largest slope $\hat{b}$ is then obtained for a stretch of 6
observations between $z=1.265$ and the largest $z=1.755$, where
one obtains a positive standardized slope of
$T(n-7,n)\approx 0.0385$ and hence $T\approx 0.0385$. This
region is shown in the right panel of Figure~\ref{fig2}. The
standardized slope is smaller than in the unconstrained test.
However, if assessing the significance of this result by 1000
simulations of the test statistic under the assumption of constant
$Y(z)$, a larger test statistic $T$ than the observed value 0.0385
was obtained for 973 runs. The result is therefore more
significant than the unconstrained one, but still comfortably
compatible with monotonicity.

\section{Conclusion}
Present-day supernova data are compatible with the monotonicity
constraint in the Hubble diagram. Since this constraint is
kinematical, {i.e.} independent of the presently debated issues
related to inflation, dark energy or dark matter, it will be
worthwhile to keep testing it as new data come in. For this
purpose, a `running gradient' type method like the one by P. Hall
and N. Heckmann (2000) \cite{hh} used here seems well adapted for
its flexibility and numerical ease of use.
\vskip 1cm \noindent
{\it Acknowledgements:}
Nicolai Meinshausen is supported by a fellowship of DFG (Deutsche Forschungsgemeinschaft);
John Rice acknowledges the support of a grant from the US National Science Foundation, AST-0507254; Thomas Sch\"ucker acknowledges a research semester granted to him by his home institutes and thanks Vaughan Jones for his generous hospitality and coaching.

\end{document}